\documentclass[twocolumn,showpacs,aps,pra,amsmath,amssymb]{revtex4-1}
\usepackage{graphicx}
\usepackage{dcolumn}
\usepackage{bm,color}
\usepackage{txfonts}
\usepackage{amsmath,epsfig}
\usepackage{epstopdf}

\newcommand{\eq}{Eq.~}
\newcommand{\eqs}{Eqs.~}
\newcommand{\fig}{Fig.~}
\newcommand{\figs}{Figs.~}
\newcommand{\refs}{Refs.~}
\newcommand{\cf} {cf.~}
\newcommand{\ie} {i.e.~}
\newcommand{\eg} {e.g.~}

\newcommand{\etal} {{\it et al}.~}

\begin{document}

\author{Francesco Ciccarello}
\affiliation{NEST, Scuola Normale Superiore and Istituto Nanoscienze-CNR, Piazza dei Cavalieri 7, I-56126 Pisa, Italy\\CNISM and Dipartimento di Fisica, 
Universita' degli Studi di Palermo, Viale delle Scienze, Ed. 18, I-90128 Palermo, Italy}
\email{francesco.ciccarello@sns.it}

\pacs{03.67.Hk, 03.67.Ud}

\title{Interaction between hopping and static spins in a discrete network}
 
\date{\today}
\begin{abstract}
We consider a process where a spin hops across a discrete network and at certain sites couples to static spins. While this setting is implementable in various scenarios (e.g quantum dots or coupled cavities) the physics of such processes is still basically unknown. Here, we take a first step along this line by scrutinizing a two-site and a three-site lattices, each with two static spins. Despite a generally complex dynamics occurs, we show a regime such that the spin dynamics is described by an effective three-spin chain. Tasks such as entanglement generation and quantum state transfer can be achieved accordingly.
\end{abstract}
\maketitle

\noindent

\section{Introduction}

A natural need in the accomplishment of quantum information processing (QIP) \cite{nc} tasks is to spread quantum information among distant qubits and/or transfer it from one to the other. Illustrative processes are entanglement \cite{nc} generation (EG) and quantum state transfer (QST) \cite{sougato, cambridge}. Unless the qubits lie close to each other (an atypical situation given that one normally wishes to retain local control over them) no direct inter-qubit interaction is usually available. This immediately brings about that a bus embodied by a third system is usually required to enable crosstalk between the qubits. This task is often accomplished by means of photons that are coupled through a Jaynes-Cummings-like interaction to each qubit, the latter being typically embodied by an atom or atom-like system (see \eg \cite{cirac,fiber,irish}). In all such cases, the number of mediating photons is usually not conserved.

Another paradigm for implementing the bus is to employ {\it spin-bearing} scattering particles \cite{imps, NJP, yuasa1, romani, habgood, prl, hida, mappaNP, daniel} such as electrons or photons, whose number is conserved during the process (in the photon case two orthogonal polarizations span an effective pseudo-spin space \cite{prl,mappaNP}). In this literature, the inter-qubit crosstalk is mediated by a {\it monochromatic} spin-bearing particle that usually undergoes one-dimensional (1D) quantum scattering from two static spins embodying the qubits. Based on the assumption of monochromaticity, one can define suitable resonance conditions \cite{NJP,yuasa1} that are analogous to those for a standard Fabry-Perot interferometer and prove that, if fulfilled, these yield an additional conserved quantity, the squared total spin of the scattering centers \cite{NJP}. By harnessing such effect a number of efficient and robust schemes for distributing entanglement between the centers have been demonstrated \cite{prl, mappaNP, daniel}.

The above picture well fits within a continuous-waveguide scenario such as a semiconductor nanowire, along which electrons and/or photons can travel and where the static spins can be implemented through magnetic impurities or quantum dots (see \eg \cite{prl,mappaNP} and references therein). Yet, in the context of QIP some of the platforms that are receiving widespread attention are discrete. This is for instance the case of cold-atom lattices \cite{sanpera} and coupled-cavity arrays \cite{reviews}. In such scenarios, it is natural to envisage processes where the static spins are located close to some sites and, importantly, the mobile mediator initially sits {\it at a given site}. This then spreads over the lattice sites in such a way that at a later time it in general acquires some non-null probability to be found at both the spins' locations. Analogously to the continuous case, one can wonder whether a process of this sort can be harnessed for efficient quantum communication purposes. Aside from the above motivations related to state-of-the-art setups, such issue is intrinsically interesting in that it involves spin-bearing mediators with well-defined position instead of well-defined momentum as those considered in \refs\cite{imps, NJP, yuasa1, romani, habgood, prl, hida, mappaNP, daniel}. 
With the exception of one study \cite{peskin} which we will discuss shortly, processes of this sort are to date essentially unexplored \cite{nota-romani}. 

The present work is intended to provide a first step towards a comprehensive knowledge of such dynamics by tackling, as our first task, the primitive setup comprising a mobile spin hopping between two sites in a way that at each site it locally couples to one of two static spins. Although within the present framework this is the simplest set-up that can be envisaged, it exhibits a non-trivial dynamical behavior, which has to our knowledge remained unaddressed to date. Nonetheless, we will show that  in the regime of strong hopping the spin dynamics decouples from the motional one and is effectively described by a simple three-spin chain.
This immediately entails that EG and QST are achievable according to the considered interaction model. We next address the case where the mobile spin can hop between three sites and show that while the above effect in general does not occur it is however exhibited when the hopping particle initially sits on the middle site. 

Despite the simplicity of the mathematical demonstration of these effects, their implications are non-trivial in many respects. In particular, from a fundamental viewpoint they mark profound differences between continuous and discrete settings (from this perspective our focus on typical QIP figures of merit is simply a tool to acquire insight into the quantum dynamics regardless of applicative facets). From an applicative viewpoint, the low dimensionality of the investigated settings can provide experimentalists with alternative strategies for achieving QIP in the imminent future (when it is believed that at a first stage only small-size quantum coherent set-ups such as two coupled cavities will be accessible in the lab).

Throughout, we will carry out our analysis so as to take account of both the cases where the spin-spin interaction is described by a Heisenberg and an $XY$-isotropic model. Such flexibility makes our theory effective in a variety of actual settings such as quantum dots and cavity-QED scenarios.

As mentioned earlier, Peskin \etal \cite{peskin} recently tackled a setting similar to those in this Letter. Aside from the different  coupling model (they deal with $XX$-type interactions) it is important to stress that unlike in \cite{peskin} here the hopping particle can couple to at most only one static spin at each site. Also, we focus on static spins that do not possess any free Hamiltonian, which rules out resonance-based effects that are crucial in the EG scheme in Ref.~\cite{peskin}.

The present paper is organized as follows. In Section \ref{2site}, we introduce the two-site setup, give the associated Hamiltonian and present some related numerical results. Based on these and other arguments, we motivate our plan to focus on the regime of strong hopping. In Section \ref{dyneff}, we calculate the effective Hamiltonian and some of its relevant features are thoroughly discussed. In Sections \ref{communication} and \ref{QST}, we use the effective Hamiltonian in the above regime to investigate the system's performances in terms of EG and QST, respectively. In Section \ref{3site}, we tackle the three-site setup and show the conditions under which a behavior analogous to the two-site case can arise. Finally, in Section \ref{conclusions}, we draw our conclusions.

\section{Two-site setup} \label{2site}

We consider two static spin-1/2 particles, labeled with 1 and 2, whose spin operators are denoted by $\hat{\mathbf{S}}_1$ and $\hat{\mathbf{S}}_2$, respectively. Although not directly interacting, the particles can crosstalk through a spin-bearing particle $e$ hopping between the two sites of a lattice as shown in \fig1. We call $|1\rangle$ ($|2\rangle$) the motional state of $e$ when it lies at site 1 (site 2), whereas its associated spin operator is denoted by $\hat{\mbox{\boldmath$\sigma$}}$. The Hamiltonian is modeled according to (here and throughout we set $\hbar\!=\!1$)
\begin{equation}\label{H}
\hat{H}=\hat{H}_{\rm hop}+\hat{V}\,\,,
\end{equation}
where
\begin{eqnarray}
\hat{H}_{\rm hop}&=&\eta \left(|1\rangle\langle2|+|2\rangle\langle1|\right)\,\,,\label{H-hop}\\
\hat{V}&=&\sum_{x=1}^2|x\rangle\langle x|  \left\{J_{XY}\left(\hat{\sigma}_{+}\hat{S}_{x-}\!+\!\hat{\sigma}_{-}\hat{S}_{x+}\right)\!+\!J_{z}\,\hat{\sigma}_{z}\hat{S}_{xz}\right\}\label{V}\,\,.
\end{eqnarray}
In \eq(\ref{H}), $\hat{H}_{\rm hop}$ and $\hat{V}$ are the kinetic and interaction Hamiltonians, respectively. In \eq(\ref{V}), the spin-spin coupling between $e$ and each static spin within braces consists of an $XY$-isotropic and an Ising term with associated coupling strengths $J_{XY}$ and $J_z$, respectively. For $J_{z}\!=\!J\!=\!2J_{XY}$ a Heisenberg interaction is obtained in such a way that the factor between braces in \eq(\ref{H}) reduces to $\hat{\mbox{\boldmath$\sigma$}}\cdot\mathbf{S}_x$. When $J_z\!=\!0$ a pure $XY$-isotropic model arises.
\begin{figure}
\includegraphics[width=0.28\textwidth]{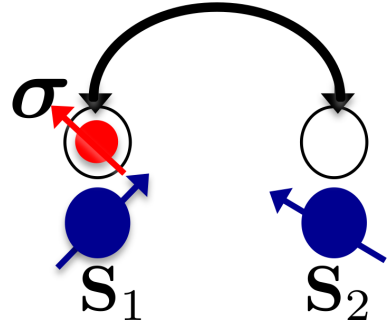}
\caption{(Color online) Sketch of the two-site setup. A mobile particle (in red) hops between two sites. At site 1 (site 2), its spin $\mbox{\boldmath$\sigma$}$ couples to spin $\mathbf{S}_1$ ($\mathbf{S}_2$). Static spins $\mathbf{S}_1$ and $\mathbf{S}_2$ are not directly interacting with each other. \label{fig1}}
\end{figure}

To begin with, we address the $XY$-isotropic case and consider the initial state $|\psi(0)\rangle\!=\!|{x\!=\!1\rangle}|{\uparrow}\rangle_e|{\downarrow\downarrow}\rangle_{12}$. At a time $t\!>\!0$ the system has evolved into the state $\rho(t)\!=\!|\Psi(t)\rangle\!\langle\Psi(t)|$, where $|\Psi(t)\rangle\!=\!\hat{U}(t)|\psi(0)\rangle$ and $\hat{U}(t)\!=\!e^{-i\hat{H}t}$. The probability to find $e$ at $x\!=\!1$ is calculated as $P_1\!=\!{\rm Tr} \,[\rho|1\rangle\!\langle1|]$ and likewise $e$ is measured in $|{\uparrow}\rangle$ with probability $P_\uparrow\!=\!{\rm Tr} \,[\rho|{\uparrow}\rangle_e\langle{\uparrow}|]$.  The density operator that describes the state of 1-2 $\rho_{12}$ is obtained upon trace of  $\rho$ over the motional and spin degrees of freedom of $e$. The overlap of $\rho_{12}$ with $|{\Psi^\pm}\rangle_{12}\!=\!(|{\uparrow\downarrow}\rangle_{12}\!\pm\!|{\downarrow\uparrow}\rangle_{12})\!/\!\sqrt{2}$, \ie respectively the maximally entangled triplet and singlet states of 1-2, can be measured through the fidelity $F_{\pm}\!=\!\langle{\Psi^{\pm}}|\rho_{12}|{\Psi^{\pm}}\rangle$. In \fig2, we consider the three different settings $\eta/J\!=\!1, 2 ,10$. For each of these, we plot against time $P_1$ (left figure) along with $F_{\pm}$, $P_{\uparrow}$ and $E$ (right figure), where $E$ is the logarithmic negativity \cite{logneg} of $\rho_{12}$ (at this stage all the results are numerical). In the case of competitive hopping and spin-spin interactions (first-row plots) quite a complex behavior is exhibited. While the mobile-particle hopping between the two sites is not harmonic (as in the absence of spins 1 and 2), the probability to find $e$ in $|{\uparrow}\rangle$ roughly fluctuates between 0 and values higher than $\!\simeq\!0.6$. Notice that, on a rough approximation, $F_+$ is low (high) when $P_{\uparrow}$ is high (low) and that the entanglement amount mostly behaves as $F_+$. Remarkably, a significant overlap with the singlet $|{\Psi^-}\rangle_{12}$ is developed given that $F_-$ can exceed 0.3. When the hopping rate $\eta$ is twice as large as the spin-spin interaction strength $J$ (middle-row plots) a more regular behavior arises with the time evolution of $P_1$, $P_\uparrow$, $F_+$ and $E$ that now closely resembles an oscillatory function. In this regime, the singlet fraction is significantly reduced compared to the previous case. Finally, in the strong-hopping regime $\eta\!\gg\! J$ (lower-row plots) a regular and harmonic behavior is exhibited. The mobile particle harmonically hops between 1 and 2, in fact as if the static spins were absent. Both $P_\uparrow$ and $F_+$ oscillate between 0 and 1 with $P_\uparrow$ taking value 0 (1) when $F_+\!=\!1$ ($F_+\!=\!0$). On the other hand $F_-\!\simeq\!0$, which witnesses that the dynamics takes place entirely within the triplet subspace. The behavior of 1-2 entanglement $E$ essentially follows the triplet fidelity becoming maximum whenever $F_+\!=\!1$.
\begin{figure*}
\includegraphics[width=0.48\textwidth]{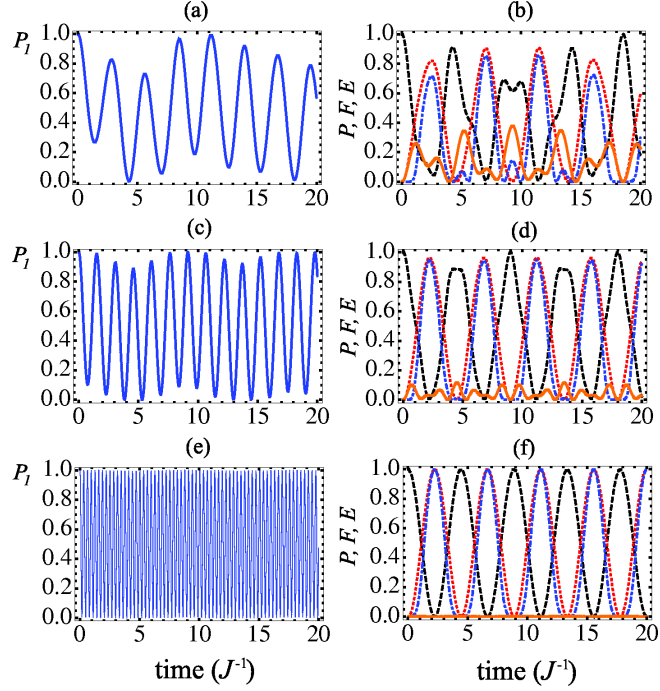}
\caption{(Color online) (a), (c), (e) Probability to find $e$ at site $x\!=\!1$ $P_1$ against time for $\eta/J\!=\!1$ (a), $\eta/J\!=\!2$ (c) and  $\eta/J\!=\!10$ (e). (b), (d), (f) Time evolution of: the fidelity of $\rho_{12}$ with respect to $|\Psi^-\rangle_{12}$ (orange solid line) and $|\Psi^+\rangle_{12}$ (red dotted), the logarithmic negativity of $\rho_{12}$ (blue dot-dashed) and the probability to find $e$ in $|{\uparrow}\rangle$ $P_\uparrow$ (black dashed line) for $\eta/J\!=\!1$ (b), $\eta/J\!=\!2$ (d) and  $\eta/J\!=\!10$ (f). Figures on the same row correspond to the same set value of $\eta/J$. The system's initial state is $|\Psi\rangle\!=\!|x\!=\!1,{\uparrow}\rangle_e|{\downarrow\downarrow}\rangle_{12}$.
All the plotted curves were obtained numerically by assuming an $XY$-isotropic spin-spin coupling, \ie $J_z\!=\!0$, $J_{XY}\!=\!J$ .}
\end{figure*}

In the light of such outcomes \cite{nota-heis}, we henceforth focus on the strong-hopping regime $\eta\!\gg\!\{J_{XY},J_z\}$. Indeed, while in this case the dynamics appears simple enough to be tackled analytically, the above-illustrated emergence of $|\Psi_+\rangle$ starting from the assumed initial spin state closely resembles the continuous analogue \cite{NJP} of the present setting \cite{abuse}. On the other hand, in the continuous case (CC) the eigenvalues of the kinetic Hamiltonian of $e$ coincide with the full energy spectrum \cite{NJP} (because there the interaction between $e$ and each static spin occurs through contact potentials \cite{NJP,yuasa1}). This makes the strong-hopping regime the natural one to scrutinize in the present scenario for developing a comparison between the discrete and continuous case, \ie a major goal of this work. Finally, the emergence of {\it maximally} entangled states during the evolution, differently from \figs2(b) and 2(d), makes this regime especially attractive for the sake of QIP tasks.

\section{Effective dynamics} \label{dyneff}

To acquire analytical insight into the behaviors in \figs2(e) and (f), we start observing that in the motional Hilbert space the hopping Hamiltonian is straightforwardly diagonalized as
\begin{equation}\label{H-hop-diag}
\hat{H}_{\rm hop}=\eta\left(|\varphi_+\rangle\langle\varphi_+|-|\varphi_-\rangle\langle\varphi_-|\right)\,\,,
\end{equation}
where 
\begin{equation} \label{phipm}
|\varphi_{\pm}\rangle=\frac{|1\rangle\pm|2\rangle}{\sqrt{2}}\,\,.
\end{equation}
In terms of states $|\varphi_{\pm}\rangle$'s, the site states are decomposed as $|1\rangle\!=\!(|\varphi_{+}\rangle\!+\!|\varphi_{-}\rangle)/\!\sqrt{2}$ and $|2\rangle\!=\!(|\varphi_{+}\rangle\!-\!|\varphi_{-}\rangle)/\!\sqrt{2}$.

In the interaction picture, by taking $\hat{H}_0\!=\!\hat{H}_{\rm hop}$ as the free Hamiltonian, the projectors $\{|x\rangle\!\langle x|\}$ ($x\!=\!1,2$) evolve as
\begin{eqnarray}\label{11t}
|1\rangle\langle 1|^{(I)}(t)&=&\frac{\openone_{{\rm mot}}}{2}+\left(|\varphi_+\rangle\langle \varphi_-|\,e^{2i\eta t}+{\rm h.c.}\right)\,\,,\,\,\,\,\,\,\,\\\
|2\rangle\langle 2|^{(I)}(t)&=&\frac{\openone_{{\rm mot}}}{2}-\left(|\varphi_+\rangle\langle \varphi_-|\,e^{2i\eta t}+{\rm h.c.}\right)\,\,,\,\,\,\,\,\,\,\label{22t}
\end{eqnarray}
where $\openone_{{\rm mot}}\!=\!|\varphi_+\rangle\langle \varphi_+|+|\varphi_-\rangle\langle \varphi_-|$ is the identity operator in the motional Hilbert space.

Upon use of \eqs(\ref{V}), (\ref{11t}) and (\ref{22t}) in the interaction picture the interaction Hamiltonian reads
\begin{eqnarray}\label{V-int}
\hat{V}^{(I)}(t)&=&\frac{J_{XY}}{2}\, \left[\hat{\sigma}_{+}\,(\hat{S}_{1-}\!+\!\hat{S}_{2-})+{\rm h.c.}\right]+\frac{J_{z}}{2}\, \,\hat{\sigma}_z\,(\hat{S}_{1z}\!+\!\hat{S}_{2z})\nonumber\\
&+&\left(|\varphi_+\rangle\langle \varphi_-|\,e^{i2\eta t}+{\rm h.c.}\right)\,\left\{\frac{J_{XY}}{2}\,\hat{\sigma}_{+}\,(\hat{S}_{1-}\!-\!\hat{S}_{2-})+{\rm h.c.}\right\}\nonumber\\
&+&\left(|\varphi_+\rangle\langle \varphi_-|\,e^{i2\eta t}+{\rm h.c.}\right)\,\frac{J_{z}}{2}\,\hat{\sigma}_{z}\,(\hat{S}_{1z}\!-\!\hat{S}_{2z})\,\,.
\end{eqnarray}
Evidently, (\ref{V-int}) comprises a constant and a time-dependent part. The former thus survives in the limit of strong hopping, \ie $\eta\!\gg\! \{J_{XY}, J_z\}$, even to the first order in the coupling strengths. This marks a major difference from the case of a hopping photon, free of internal degrees of freedom, coupled to two atoms through a Jaynes-Cummings-like interaction \cite{irish}, where first-order contributions to the interaction Hamiltonian vanish. This is essentially due to the fact that the interaction Hamiltonian (\ref{V}) is quadratic in the degrees of freedom of $e$ \cite{nota-quadratic}. 

A pivotal feature of (\ref{V-int}) is that the time-independent term couples the $e$'s spin to the {\it total spin} of particles 1 and 2, \ie $\hat{\mathbf{S}}_{12}\!=\!\hat{\mathbf{S}}_1\!+\!\hat{\mathbf{S}}_2$. So does not the time-dependent one (where the difference of such spins $\hat{\mathbf{S}}_1\!-\!\hat{\mathbf{S}}_2$ is involved). In the strong-hopping limit $\eta\!\gg\! \{J_{XY}, J_z\}$ here considered, up to the first order in the coupling strengths all of the rotating terms in \eq(\ref{V-int}) proportional to $e^{\pm i 2\eta t}$ negligibly contribute to the effective dynamics. Thereby in the light of \eqs(\ref{H}), (\ref{H-hop}), (\ref{H-hop-diag}) and (\ref{V-int}) the effective Hamiltonian in the Schr\"odinger picture simply reduces to
\begin{equation}\label{Heff}
\hat{H}_{\rm eff}\!\simeq \eta \left(|1\rangle\langle2|+|2\rangle\langle1|\right)+\hat{V}_{\rm eff}\,\,,
\end{equation}
where 
\begin{equation}
\hat{V}_{\rm eff}=\frac{J_{XY}}{2}\left(\hat{\sigma}_{+}\hat{S}_{12-}\!+\!\hat{\sigma}_{-}\hat{S}_{12+}\right)\!+\!\frac{J_z}{2} \hat{\sigma}_z \hat{S}_{12z}\label{Veff}
\end{equation}
depends only on spin variables, at variance with (\ref{V}).
\begin{figure}
\includegraphics[width=0.28\textwidth]{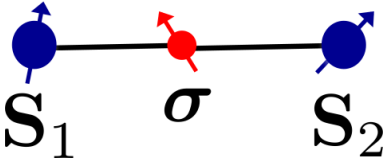}
\caption{(Color online) Effective three-site spin chain describing the spin dynamics in the strong-hopping regime. The coupling strengths between $e$ and 1 (2) are $J_{XY}/2$ and $J_z/2.$
\label{fig3}}
\end{figure}
In spite of their straightforward derivation, \eqs(\ref{Heff}) and (\ref{Veff}) deserve several comments. In the strong-hopping regime, the motional and spin degrees of freedom turn out to be effectively {\it decoupled}. Importantly, the spin term of (\ref{Veff}) in fact coincides with the Hamiltonian of an open three-spin chain, where the spin of $e$ plays the role of the central spin symmetrically coupled with equal strengths $J_{XY}/2$ and  $J_{z}/2$ to the end spins 1 and 2, as sketched in \fig3. Notice, though, that the coupling strengths are halved. The above can be given an intuitive physical interpretation: When $e$ hops between 1 and 2 much faster than the rate at which it interacts with them, it behaves as a stationary spin lying at their locations at the same time. Particle $e$ therefore sees spins 1 and 2 collectively. Due to the structure of the interaction Hamiltonian, this entails that the squared total spin of 1 and 2 $\hat{\mathbf{S}}_{12}^2$ is a conserved quantity, namely 
\begin{equation}\label{conservation}
[\hat{H}_{\rm eff}, \hat{\mathbf{S}}_{12}^2]=0\,\,,
\end{equation} 
as it is immediate to see from \eqs(\ref{Heff}) and (\ref{Veff}). This is already enough to explain why in \fig2(f) the overlap with the singlet $|\Psi^-\rangle_{12}$ is identically zero: the initial spin state $|{\uparrow}\rangle_e|{\downarrow\downarrow}\rangle_{12}$ fully lies within the triplet subspace. Due to \eq(\ref{conservation}) and conservation of $\hat{\sigma}_z\!+\!\hat{S}_{12z}$ (projection of the total spin along the $z$-axis), the system's state at any later time necessarily is a superposition of $|{\uparrow}\rangle_e|{\downarrow\downarrow}\rangle_{12}$ and $|{\downarrow}\rangle_e|{\Psi^+}\rangle_{12}$.

Such effective conservation of $\hat{\mathbf{S}}_{12}^2$ shares features with the CC in the regime where a monochromatic mobile particle of wave vector $k$ propagating along a 1D wire scatters from the static spins under the resonance conditions (RCs) $k x_0\!=\!n\pi$ \cite{NJP,mappaNP} ($x_0$ is the distance between 1 and 2 while $n\!\in\! \mathbb{Z}$). There, one can show \cite{NJP,mappaNP} that due to such RCs the static spins behave as if they were at the same place, which brings about an effective coupling between their total spin and the spin of the mobile particle similarly to \eq(\ref{Veff}) and thereby conservation of $\hat{\mathbf{S}}_{12}^2$. As a peculiar feature of the present set-up, though, here the occurrence of this type of effect does not require any constraint over the motional state of $e$ since to derive \eq(\ref{Heff}) we have only used $\eta\!\gg\! \{J_{XY}, J_z\}$. In particular, this takes place even if $e$ is initially in an asymmetric state such as $|x\!=\!1\rangle$ considered in \fig2, quite differently from the continuous case (CC) where the wave vector of $e$ must fulfill RCs, a condition under which the static spins are seen symmetrically. A further remarkable difference is that while in the CC \cite{NJP,mappaNP}  the effective spin-spin scattering potential under RCs is proportional to (\ref{Veff}) it nevertheless contains an extra factor given by a function $\delta(x)$ ($x$ is the continuous coordinate of the mobile particle) \cite{NJP,mappaNP}. Hence, the spin dynamics remains anyway coupled to the motional one differently from the present case [see \eqs(\ref{Heff}) and (\ref{Veff})]. It is also worth remarking that the effect behind \eqs(\ref{Heff}) and (\ref{Veff}) is in some respects reminiscent of a phenomenon studied in Ref.~\cite{dalibard}, where it was shown that a spinless particle hopping in a double quantum dot can behave as an effective {\it static}  double potential barrier able to give rise to Fabry-Perot-like effects.

\section{entanglement generation} \label{communication}

The form of (\ref{Veff}) straightforwardly yields that the system can develop a non-vanishing overlap with the maximally entangled triplet state of the static spins $|{\Psi^+}\rangle_{12}$, similarly to the CC scattering scenario \cite{NJP}. To show this, we notice that $\hat{V}_{\rm eff}$ can be block-diagonalized in the 8-dimensional overall spin space, where each block corresponds to an eigenspace of the two conserved quantities $\hat{\sigma}_z\!+\!\hat{S}_{12\,z}$ and $\hat{\mathbf{S}}_{12}^2$. Four of such eigenspaces are one-dimensional and given by $\{|{\uparrow}\rangle_e|{\uparrow\uparrow}\rangle_{12}, |{\downarrow}\rangle_e|{\downarrow\downarrow}\rangle_{12}\}$, both having eigenvalue $J_z/4$, and $\{|{\uparrow}\rangle_e|{\Psi^-}\rangle_{12}, |{\downarrow}\rangle_e|{\Psi^-}\rangle_{12}\}$ both having zero eigenvalue due to $\hat{\mathbf{S}}_{12}|{\Psi^-}_{12}\rangle\!=\!0$.

The two remaining doublets are spanned by $\{|{\uparrow}\rangle_e|{\downarrow\downarrow}\rangle_{12},|{\downarrow}\rangle_e|{\Psi^+}\rangle_{12}\}$ and $\{|{\downarrow}\rangle_e|{\uparrow\uparrow}\rangle_{12},|{\uparrow}\rangle_e|{\Psi^+}\rangle_{12}\}$). Within the one spanned by $\{|{\uparrow}\rangle_e|{\downarrow\downarrow}\rangle_{12},\, |{\downarrow}\rangle_e|{\Psi^+}\rangle_{12}\}$ (the other case is tackled analogously with due replacements) the matrix representation of $\hat{V}_{\rm eff}$ reads
\begin{equation} \label{matrix}
 \mathbf{V}_{\rm eff}\! =\!\left( \begin{array}{cc}
-J_z/4&           J_{XY}/\sqrt{2}         \\
            J_{XY}/\sqrt{2}          &  0\\
 \end{array} \right)\,\,.
 \end{equation}
In the case of the $XY$-isotropic coupling \cite{yuasa3}, \ie $J_z\!=\!0$ and $J_{XY}\!=\!J$, the eigenstates of (\ref{matrix}) are found as
\begin{equation}\label{chi1}
|{\chi_{\pm}}\rangle\!=\!\frac{|{\uparrow}\rangle_e|{\downarrow\downarrow}\rangle_{12}\pm|{\downarrow}\rangle_e|{\Psi^+}\rangle_{12}}{\sqrt{2}}\,\,,
\end{equation}
with corresponding eigenvalues
\begin{equation}\label{eps}
\varepsilon_{\pm}=\pm \frac{J}{\sqrt{2}}\,\,.
\end{equation}
Therefore, when the spin state $|{\uparrow}\rangle_e|{\downarrow\downarrow}\rangle_{12}\!=\!(|{\chi_+}\rangle\!+\!|{\chi_-}\rangle)/\!\sqrt{2}$ is prepared at time $t\!=\!0$, as in the cases in \figs2(e) and (f), at a later time $t$ the system is in the state
\begin{eqnarray}\label{Psit}
|{\psi(t)}\rangle&=&\frac{e^{-i \varepsilon_+ t}|{\chi_+}\rangle +e^{-i \varepsilon_- t}\|{\chi_-}\rangle}{\sqrt{2}}\nonumber\\&=&\cos{\left[\frac{J}{\sqrt{2}} t\right]}\,|{\uparrow}\rangle_e|{\downarrow\downarrow}\rangle_{12}+i\sin\,\left[\!\frac{J}{\sqrt{2}}t\right]\,|{\downarrow}\rangle_e|{\Psi^+}\rangle_{12}\,.\,\,\,\,\,\,\,\,\,\,\,\,\,
\end{eqnarray}
As shown by \eq(\ref{Psit}), the system oscillates between states $|{\uparrow}\rangle_e|{\downarrow\downarrow}\rangle_{12}$ and $|{\downarrow}\rangle_e|\Psi^+\rangle_{12}$ with period $2\!\sqrt{2}\pi/J$, which fully explains all the outcomes in \fig2(f).

In the case of the Heisenberg coupling, \ie $2J_{XY}\!=\!J_z\!=\!J$, the eigenstates of (\ref{matrix}) are calculated as \cite{clebsch}
\begin{eqnarray}\label{chi2}
|{\tilde{\chi}_{+}}\rangle\!=\!\frac{1}{\sqrt{3}}\,|{\uparrow}\rangle_e|{\downarrow\downarrow}\rangle_{12}+\sqrt{\frac{2}{3}}\,\,|{\downarrow}\rangle_e|{\Psi^+}\rangle_{12}\,\,,\\
|{\tilde{\chi}_{-}}\rangle\!=\!  \sqrt{\frac{2}{3}}\,|{\uparrow}\rangle_e|{\downarrow\downarrow}\rangle_{12}-\frac{1}{\sqrt{3}} \,\,|{\downarrow}\rangle_e|{\Psi^+}\rangle_{12}\,\,,
\end{eqnarray}
with corresponding eigenvalues
\begin{eqnarray}\label{epstilde}
\tilde{\varepsilon}_{\pm}=\frac{-1\pm3}{8}\,J\,\,.
\end{eqnarray}
Hence, when the spin state $|{\uparrow}\rangle_e|{\downarrow\downarrow}\rangle_{12}\!=\!(|{\tilde{\chi}_+}\rangle\!+\!\sqrt{2}\,|{\tilde{\chi}_-}\rangle)/\!\sqrt{3}$ is prepared at time $t\!=\!0$, at a later time $t$ the system is in the state
\begin{eqnarray}\label{Psit-tilde}
|{\tilde{\psi}(t)}\rangle&=&\frac{1}{\sqrt{3}}\left\{e^{-i \tilde{\varepsilon}_+ t}|{\tilde{\chi}_+}\rangle +\sqrt{2}\,e^{-i \tilde{\varepsilon}_- t}|{\tilde{\chi}_-}\rangle\right\}\nonumber\\&\!=\!&\alpha_\uparrow(t)\,|{\uparrow}\rangle|{\downarrow\downarrow}\rangle+\alpha_\downarrow(t)\,|{\downarrow}\rangle|{\Psi^+}\rangle\,\,, 
\end{eqnarray}
where
\begin{eqnarray}
\alpha_\uparrow(t)&=&\cos{\left[\frac{3J}{8} t\right]}+\frac{1}{3}\sin{\left[\frac{3J}{8} t\right]}\,\,,\label{alphaup}\\
\alpha_\downarrow(t)&=&-\frac{2\sqrt{2}}{3}\,i\sin\,\left[\frac{3J}{8}t\right]\,\,.\label{alphadown}
\end{eqnarray}
\begin{figure}
\includegraphics[width=0.48\textwidth]{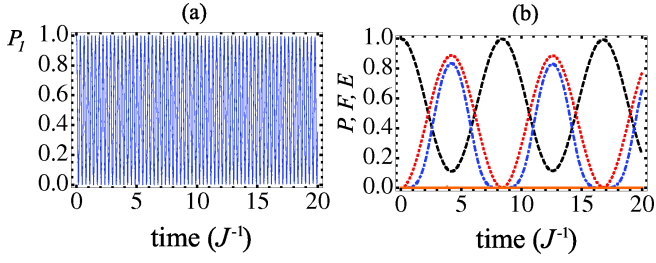}
\caption{(Color online) {\bf (a)} Probability to find $e$ at site $x\!=\!1$ $P_1$ against time. {\bf (b)} Time evolution of: the fidelity of $\rho_{12}$ with respect to $|\Psi^-\rangle_{12}$ (orange solid line) and $|\Psi^+\rangle_{12}$ (red dotted), the logarithmic negativity of $\rho_{12}$ (blue dot-dashed) and the probability to find $e$ in $|{\uparrow}\rangle$ $P_\uparrow$ (black dashed line). The system's initial state is $|\psi(0)\rangle\!=\!|x\!=\!1,{\uparrow}\rangle_e|{\downarrow\downarrow}\rangle_{12}$. All the plotted curves were obtained numerically by setting $\eta/J\!=\!10$ and assuming a Heisenberg spin-spin coupling.}
\end{figure}
\eq(\ref{Psit-tilde}) shows that an oscillatory behavior between states $|{\uparrow}\rangle_e|{\downarrow\downarrow}\rangle_{12}$ and $|{\downarrow}\rangle_e|\Psi^+\rangle_{12}$ is exhibited with period $16\pi/(3J)$. Unlike the $XY$-isotropic model [\cf\eq(\ref{Psit})], however, not the entire initial population of $|{\uparrow}\rangle_e|{\downarrow\downarrow}\rangle_{12}$ is transferred to state $|{\downarrow}\rangle_e|\Psi^+\rangle_{12}$, the maximum fidelity with respect to the latter being $F_-\!=\!8/9\simeq0.89$ according to \eq(\ref{alphadown}). In \figs4(a) and (b), we plot the same quantities in the strong-hopping regime as in \figs2(e) and (f), respectively, but under the assumption of a Heisenberg coupling. While \fig4(a) is fully analogous to  \fig2(e) in accordance with \eq(\ref{Heff}), the plots in \fig4(b), which were obtained numerically, are in excellent agreement with \eqs(\ref{Psit-tilde})-(\ref{alphadown}).

\section{Quantum state transfer} \label{QST}

As discussed in the Introduction, a question that is naturally raised is whether the setting under investigation is suitable for achieving QST \cite{sougato, cambridge} between the two static spins. To address this issue, we consider the initial state $|{\psi(0)}\rangle\!=\!|x\!=\!1\rangle|{\downarrow}\rangle_e|{\uparrow\downarrow}\rangle_{12}$ with the aim to assess whether at a later time the $\uparrow$-excitation has fully transferred from spin 1 to spin 2, \ie the condition $F_{2}\!=\!_{12}\!\langle{\downarrow\uparrow}|\rho_{12}|{\downarrow\uparrow}\rangle_{12}\!=\!1$ is fulfilled. In \fig5, we focus on the strong-hopping regime, \ie $\eta\!\gg\!\{J_{XY},J_{z}\}$, for the $XY$-isotropic (a) and Heisenberg (b) interaction models and plot $F_{2}$ as a function of time.
\begin{figure}
\includegraphics[width=0.48\textwidth]{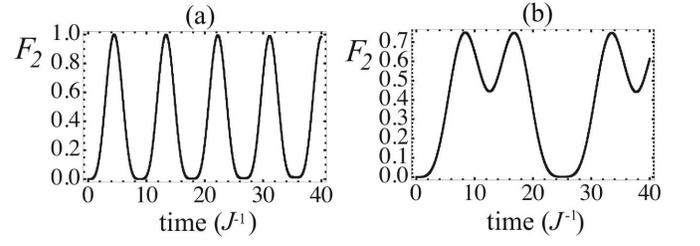}
\caption{{\bf (a)} $F_{2}$ vs. time (in units of $J^{-1}$) \label{fig5} for the $XY$-isotropic model with $\eta/J=20$. {\bf (b)} $F_{2}$ vs. time (in units of $J^{-1}$) \label{fig5} for the Heisenberg model with $\eta/J=20$. The initial state in both case is $|{\psi(0)}\rangle\!=\!|x\!=\!1\rangle|{\downarrow}\rangle_e|{\uparrow\downarrow}\rangle_{12}$.}
\end{figure}
Evidently, QST is achieved perfectly in the $XY$-isotropic case (where the condition $F_2\!=\!1$ periodically occurs) but partially for the Heisenberg model (where $F_2$ never exceed $0.75$). These outcomes straightforwardly follow from the discussed three-spin-chain effective dynamics [\cf\eq(\ref{Veff}) and \fig3] and the well-known QST performances of $XY$-isotropic \cite{cambridge} and Heisenberg-type spin chains \cite{sougato} (therefore here we do not carry out a detailed analysis).

\section{Three-site setup} \label{3site}

The arguments developed so far have shown that the two-site set-up in \fig1 can behave as an effective closed three-spin chain regardless of the initial motional state of the hopping particle. Having clarified this, it is natural to ask now whether such an interesting regime allowing for EG and QST can occur when additional sites are added to the lattice along which $e$ can hop. Notice that in such a way the distance between the static spins can be made larger. In the present Section, we tackle a three-site set-up, which is obtained from the one in \fig1 through addition of a middle site as illustrated in \fig6.
The motional Hilbert space associated with the mediator $e$ now becomes three-dimensional and is spanned by the site states $|x\!=\!1\rangle$, $|x\!=\!0\rangle$ and $|x\!=\!2\rangle$ (see \fig6), where for consistency of notation with the previous case we have labelled the middle site with $x\!=\!0$ and the left (right) one with $x\!=\!1$ ($x\!=\!2$). The Hamiltonian reads $\hat{H}_3=\hat{H}_{\rm hop3}\!+\!\hat{V}_3$, where $\hat{V}_3$ has the same form as in \eq(\ref{V}) (apart from acting on a larger Hilbert space) while the kinetic Hamiltonian $\hat{H}_{\rm hop3}$ now becomes
\begin{eqnarray}
\hat{H}_{\rm hop3}&=&\eta \left(|1\rangle\langle0|+|0\rangle\langle2|+{\rm h.c.}\right)\,\,.\label{H-hop3}
\end{eqnarray}
In the motional Hilbert space, $\hat{H}_{\rm hop}^{(3)}$ has the eigenstates
\begin{eqnarray}
|\varphi_{\pm}\rangle&=&\frac{1}{2}\,|1\rangle\pm \frac{1}{\sqrt{2}}\,|0\rangle+\frac{1}{2}\,|2\rangle\,\,,\label{H-hop-eigen-1}\\
|\varphi_0\rangle&=&\frac{|1\rangle-|2\rangle}{\sqrt{2}}\label{H-hop-eigen-2}\,\,
\end{eqnarray}
with corresponding eigenvalues $\pm\eta$ and $0$, respectively.
\begin{figure}
\includegraphics[width=0.28\textwidth]{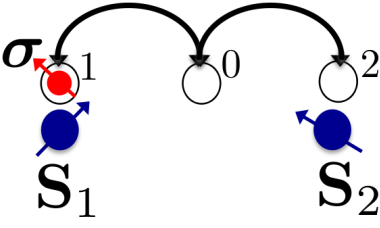}
\caption{(Color online) Sketch of the three-site setup. A mobile particle (in red) hops between three sites, labelled with 1, 0 and 2 (from left to right). At site 1 (site 2), its spin $\mbox{\boldmath$\sigma$}$ couples to spin $\mathbf{S}_1$ ($\mathbf{S}_2$). Static spins $\mathbf{S}_1$ and $\mathbf{S}_2$ are not directly interacting with each other. \label{fig2}}
\end{figure}

In terms of states (\ref{H-hop-eigen-1}) and (\ref{H-hop-eigen-2}), the site states $\{|x\rangle\}$ are expanded as
\begin{eqnarray}
|1\rangle&=&\frac{|\varphi_+\rangle-|\varphi_-\rangle}{\sqrt{2}}+\!\frac{1}{\sqrt{2}}|\varphi_0\rangle\,\,,\label{site1}\\
|0\rangle&=&\frac{|\varphi_+\rangle-|\varphi_-\rangle}{\sqrt{2}}\label{site0}\,\,,\\
|2\rangle&=&\frac{|\varphi_+\rangle-|\varphi_-\rangle}{2}-\!\frac{1}{\sqrt{2}}|\varphi_0\rangle\label{site2}\,\,.
\end{eqnarray}
Here again, we focus on the regime of strong coupling $\eta\!\gg\!\{J_{XY},J_z\}$. Thereby, the expansion of projectors $|1\rangle\langle1|$ and $|2\rangle\langle2|$ through use of \eqs(\ref{site1}) and (\ref{site2}), respectively, is approximated as
\begin{eqnarray}\label{xeff}
|{x}\rangle\langle {x}|&\simeq \frac{1}{4}\left(|\varphi_+\rangle\langle \varphi_+|+ |\varphi_-\rangle\langle \varphi_-|\right)+\frac{1}{2}|\varphi_0\rangle\langle\varphi_0|\,\,\,\,\,\,(\forall x=1,2)\,.\,\,\,\,\,\,\,\,
\end{eqnarray}
Similarly to the reasoning leading to \eq(\ref{Veff}), to derive \eq(\ref{xeff}) we have neglected terms rotating (in the interaction picture) as $e^{\pm2 i \eta t}$ like $|\varphi_\pm\rangle\langle \varphi_\mp|$ and as $e^{\pm  i \eta t}$ like $|\varphi_\pm\rangle\langle \varphi_0|$ and ${\rm h.c.}$ Hence, as shown by \eq(\ref{xeff}) the effective representations of $|1\rangle\langle1|$ and  $|2\rangle\langle2|$ do coincide in the strong-hopping regime, in such respect  analogously to the 2-site case [\cf\eqs(\ref{11t}) and (\ref{22t}) when the rotating terms are negligible]. It is immediate to see that this entails that the interaction Hamiltonian $\hat{V}_3$ takes the effective form
\begin{equation}
\hat{V}_{3{\rm eff}}=|1\rangle\langle1|\left[J_{XY}\left(\hat{\sigma}_{+}\hat{S}_{12-}\!+\!\hat{\sigma}_{-}\hat{S}_{12+}\right)\!+\!J_z \hat{\sigma}_z \hat{S}_{12z}\right].\label{Veff3}
\end{equation}
Evidently, according to \eq(\ref{Veff3}) $\mathbf{S}_{12}^2$ is a conserved quantity as in the 2-site case [\cf \eq(\ref{Veff})]. Here, however, the strong-hopping-induced decoupling between the motional and spin dynamics in general {\it does not} take place. Rather, the picture is more similar to the CC scenario under RCs \cite{NJP, mappaNP}, where one ends up with a single spin-dependent $\delta$-like barrier with a spin factor analogous to that in \eq(\ref{Veff3}) [the operator $|1\rangle\langle1|$ here plays the role of $\delta(x\!-\!x_1)$ in the CC where $x_1$ is the continuous coordinate of spin 1]. In the light of \eq(\ref{Veff3}), the spin dynamics thus in general mixes with the motional one.

Despite the above picture, a judicious and realistic choice of the initial conditions can allow for the effective three-spin-chain dynamics in \fig3 to hold in the present case either. To show this, we first notice that due to \eqs(\ref{xeff}) and (\ref{Veff3}) $|\varphi_{\pm}\rangle\langle\varphi_{\pm}|$ and $|\varphi_{0}\rangle\langle\varphi_{0}|$ commute with the full Hamiltonian. Furthermore, according to \eq(\ref{site0}) the state $|x\!=\!0\rangle$ does not overlap $|\varphi_0\rangle$. Hence, if $e$ initially lies at the middle site $x\!=\!0$ the dynamics entirely takes place out of $|\varphi_0\rangle$, \ie the effective involved motional Hibert subspace is spanned by $\{|\varphi_{\pm}\rangle\}$. {\it Under such conditions}, in \eq(\ref{xeff}) one can thereby replace $|\varphi_+\rangle\langle \varphi_+|\!+\! |\varphi_-\rangle\langle \varphi_-|$ with the identity operator and set $|\varphi_0\rangle\langle \varphi_0|\!=\!0$. Hence, in \eq(\ref{Veff3}) $|1\rangle\langle 1|$ can be in fact substituted with 1/4 in such a way that the effective interaction Hamiltonian takes the form
\begin{equation}
\hat{V}_{3{\rm eff}}^{(0)}=\frac{J_{XY}}{4}\left(\hat{\sigma}_{+}\hat{S}_{12-}\!+\!\hat{\sigma}_{-}\hat{S}_{12+}\right)\!+\!\frac{J_z}{4} \hat{\sigma}_z \hat{S}_{12z}\,\,,\label{V3eff0}
\end{equation}
where the superscript ``(0)" is a reminder that the initial conditions are such that $e$ initially lies at $x\!=\!0$ (in addition to the strong-hopping assumption).
\eq(\ref{V3eff0}) is analogous to \eq(\ref{Veff}) apart from the effective coupling strengths, which now become $J_{XY}/4$ and $J_z/4$ instead of $J_{XY}/2$ and $J_z/2$, respectively. Such halving can be interpreted as due to the larger size of the lattice along which $e$ is allowed to hop, which reduces the probability to find it at a given static-spin location. Clearly, aside from such  lower rates, all of the behaviors in terms of EG and QST analyzed in Sections \ref{communication} and \ref{QST}, respectively, hold here as well since so does the effective three-spin-chain effective spin dynamics.

It is worth remarking that from an experimental viewpoint the requirement that $e$ needs to start from the middle site is an advantageous feature. This can indeed favor the task to keep $e$ well-separated from the two static spins (\eg during the preparation stage of the system's initial spin state).

\section{Conclusions} \label{conclusions}

In this paper, we have considered a set-up comprising a spin-bearing particle hopping between two static spins, with which it locally interacts, with the goal to assess whether it allows to efficiently accomplish QIP tasks such as entanglement generation and QST between the static spins. 
When the mobile particle is allowed to hop between only two sites, in spite of the complex behavior that is generally exhibited we have shown that in the strong-hopping regime the motional and spin dynamics decouple in a way that the latter is fully described by an effective three-spin chain. While this circumstance entails effective conservation of the squared total spin of the static particles, in significant analogy with the continuous-case scattering scenario, it endows the setting with the potential of the above chain to perform quantum communication tasks. Hence, perfect or high-fidelity generation of the maximally entangled triplet state of the static spins is possible depending on the interaction model. Likewise, in line with previous studies \cite{cambridge,sougato} QST can take place perfectly or not in accordance with the considered coupling model.

Next, we have tackled the case that the mobile spin hops along a three-site lattice with the two static spins lying close to its ends. We have shown that, at variance with the two-site setting, here the motional and spin dynamics in general do not decouple under strong-hopping conditions. However, if the mobile particle is initially placed at the middle site then the effective decoupling takes place with the spin dynamics once again described by a three-site spin chain, even though with lower associated interaction strengths.

It is natural to wonder whether the effects illustrated here can be extended to setups having a number of lattice sites $N$ larger than 3. Clearly, the case $N\!\gg\!1$ is ruled out since in such conditions our approximation to neglect rotating terms to derive the first-order effective interaction is no more valid given that the free-hopping-Hamiltonian spectrum tends to a continuous band. As for values of $N$ such that $N\!\ge\!4$ but low enough to make the above approximation still valid, a natural and non-trivial extension of the effects explored in this work appears problematic even for $N\!=\!4,5$. This of course does not rule out the possibility that different mechanisms and /or regimes for achieving quantum communication tasks be effective in such a broader scenario, both with linear lattices such as those tackled here and, more in general, graphs. In this respect, our work can be useful for prompting further investigations along these lines.

Our findings are prone to be tested in various sorts of actual settings. One possible implementation employs a set-up comprising single-electron quantum dots (QDs) \cite{loss}, where each involved particle is implemented through an electron \cite{peskin}. This can embody a static spin when confined within an isolated QD, while an array of tunnel-coupled QDs can enable an electron to hop between them. Electrons occupying close enough QDs overlap their respective wavefunctions in a way that the coupling between their spins occurs through an exchange, \ie Heisenberg, interaction \cite{divi}. 

A further setting where the effects highlighted in this work can be observed is a coupled-cavity array \cite{reviews}, where each end cavity sustains two orthogonal photonic polarizations and interacts with a $\Lambda$-type three-level atom. The electric-dipole selection rules are such that the transitions between one of the two ground states and the excited state occur by absorption/emission of photons with orthogonal polarizations (for further details see \refs\cite{prl,mappaNP}). If the cavity frequency is detuned from the atomic transitions, the atomic excited level is only virtually populated and the ground doublet embodies an effective spin 1/2 that couples to the photonic pseudo spin 1/2 (encoded in the polarization degree of freedom).

It should be remarked that, from a merely {\it applicative} perspective, in the QDs implementation it may be more convenient to arrange an actual three-spin chain by assembling three single-electron tunnel-coupled QDs, instead of the afore-mentioned setting. In such a case, ours can be regarded as an alternative strategy. In contrast, the above argument does not hold for the cavity-QED implementation, basically because the atoms do not exhibit any direct interaction. This makes the applications arising from our study especially promising within a coupled-cavity framework.

\begin{acknowledgments}
Fruitful discussions with  D.~Burgarth, C.~Di Franco, G.~M.~Palma, M.~Paternostro and M. Zarcone are gratefully acknowledged.
\end{acknowledgments}

\end{document}